\documentstyle[12pt,epsfig]{article}

\def \logo {\vbox to 10.5 mm{
\hbox {}
\hbox to 59 true  mm
{
\def\magps { }
\ifnum\mag=1000 \def\magps {1.2 1.2
scale } \fi
\hfil}\vfil}}

\newcommand{\ee}{$\mbox{e}^+\mbox{e}^-$}
\newcommand{\WW}{$\mbox{W}^+\mbox{W}^-$}
\newcommand{\MW}{$M_{\mbox{W}}$}
\begin{document}

\begin{titlepage}

\logo

\vspace*{1.5cm}

\begin{flushright}
MC-TH-97/04-rev\\ MAN/HEP/97/3-rev\\ June
11, 1997
\end{flushright}

\vspace*{1.5cm}

\begin{center}
\Large{\bf Estimating the effects of
Bose-Einstein correlations
on the W mass measurement at LEP2}
\end{center}

\normalsize
\begin{center}
{\bf
V.~Kartvelishvili\footnote{On leave
from High Energy Physics Institute,
Tbilisi State University, Georgia.}
} \\
Department of Physics and Astronomy,
University of Manchester, UK, \\
{\bf R.~Kvatadze\footnotemark[1]} \\
Institut de Physique Nucl{\'e}aire de 
Lyon, France \\
and \\
{\bf R.~M\o ller} \\
Niels Bohr Institute, University of
Copenhagen, Denmark.
\end{center}
\vspace*{2.0cm}

\begin{abstract}
The influence of Bose-Einstein
correlations on the determination
of the mass of the W boson in
 \ee\ $\rightarrow$ \WW\
$\rightarrow$ 4~jet
 events at LEP2 energies is studied,
using a global event weighting
method.
We find that it is possible to keep
the systematic error on the W mass
from this source below 20 MeV, if
suitable precautions are taken in the
experimental analysis.
\end{abstract}

\end{titlepage}

\section{Introduction}

The accurate determination of the mass of the W boson is expected to be 
 one of the most important standard physics results to be obtained from 
 LEP2 measurements.
 Since \MW\ is one of the key parameters of the electroweak theory,
 precise knowledge of the W mass allows one to test the
 Standard Model (SM), and together with the top quark mass
 can be used to constrain the allowed range of the Higgs boson
 mass in the SM, or restrict the parameter space
 of other ``new physics''.
 The present value of \MW\ is based on the direct
 measurements in $\bar{\mbox{p}} \mbox{p}$ interactions at CERN {\cite{tt1}}
 and at the Tevatron {\cite{tt2,tt3}}.
 The combined result from these studies
 is \MW\ = $80.33 \pm 0.15$~GeV {\cite{pdg}}.
 Data from LEP2 recorded in 1996 at 161 and 172 GeV are expected
 to lead already to a comparable error when they are
 fully analysed and the results of the four LEP experiments
 are combined.
 When all existing data from the CDF and D0 experiments
 (more than 100~pb$^{-1}$ per experiment)
 are analyzed, it is envisaged that the
 error on the W mass will be reduced to about 70~MeV.
 On the other hand, an indirect determination of
 \MW\ from a SM fit using LEP and SLC measurements give an
 error in the W mass of approximately 40~MeV (for a Higgs mass of
 300~GeV) {\cite{tt4}}, thus setting the scale for a significant new
 test of the model.

Three different methods have been proposed for the determination
 of the W mass in \ee\ annihilation
 at LEP2 {\cite{tt5,tt6,tt7}}:
\begin{itemize}
\item[-]
 The threshold cross-section measurement of the process
 \ee $\rightarrow$ \WW\ .
 It has been shown that maximum sensitivity to \MW\ is
 achieved for the energy $\sqrt{s} \simeq 161$~GeV.
 Now these measurements are complete and the error in the W mass is
 about 450~MeV for a single experiment {\cite{opalww}}, with a combined
 result $80.4 \pm 0.22$ GeV {\cite{lepww}}.
\item[-]
 The measurement of the charged lepton end-point energy.
 This gives an estimated error exceeding 300~MeV {\cite{tt5,tt8}}
 for the total integrated luminosity expected at LEP2 and is
 therefore not competitive with the other two methods.
\item[-]
 The direct reconstruction of the \MW\ from the
 final state particles. This is considered to be the most promising
 method. It relies on the use of the four constraints from the energy-
 momentum conservation. An additional constraint can be added
 by the assumption that the two W bosons have equal masses.
 Two decay channels --- $q\bar qq\bar q$
 (four jets) and $q\bar ql\nu$ (two jets plus leptons) --- can
 be used in this analysis.
 The estimated total error for the two
 channels combined is 34~MeV, assuming four experiments
 each collecting 500~${{\rm pb}}^{-1}$ data at
 $\sqrt{s} = 175$~GeV {\cite{tt7}}.
\end{itemize}

For the process 
 \ee $\rightarrow$ \WW\
 at LEP2 energies, the typical separation of the
 two decay vertices of Ws in space
 and time is of order of 0.1~fm, much smaller than
 the hadronization scale ($\approx 0.5$~fm).
 Thus, when both Ws decay hadronically, the hadronization regions 
 of the W$^+$ and W$^-$ overlap, and colour interconnection \cite{tt9,tt0}
 during the hadronization as well as
 Bose-Einstein (BE) correlations {\cite{tt10}}
 between identical bosons among the decay products of the two Ws
 can couple the two systems and thereby affect the W mass measurement.

At present it is not excluded that these effects
 could each contribute about 100~MeV to the theoretical
 uncertainty, which would make the four jet channel essentially
 useless for the W mass measurement.
 Hence, apart from the intrinsic interest in these phenomena
 the studies of colour interconnection and
 Bose-Einstein correlations are very important, and reliable
 estimates of the size of their effect on the reconstructed W mass
 are highly desirable in order to evaluate their contributions to the
 systematic error in this measurement.

The first attempt to investigate the influence of
 Bose-Einstein correlations in the determination of W mass
 was made in {\cite{tt10}}, where the LUBOEI algorithm,
 implemented in the JETSET Monte-Carlo program {\cite{tt11}},
 was used. The basic assumptions of this approach are that
 BE effects are local in phase space and do not alter such
 characteristics as the event multiplicity or the cross section.
 The momenta of the bosons produced at the
 hadronization stage are shifted by amounts calculated to 
 reproduce the
 two particle correlation function expected for a
 source with a gaussian space-time distribution.
 This procedure, however, does not preserve momentum
 and energy simultaneously.
 The violation is not large, however, and energy
 conservation is restored in an ad hoc
 way by rescaling all particle momenta with a common
 factor. The latter procedure introduces an artificial
 shift in the W mass, even if there are
 no Bose-Einstein correlations between the
 particles coming from different Ws.
 The procedure also tends to make the jets more narrow,
 thereby decreasing the individual jet masses.
 Assuming a source radius of 0.5 fm,
 the corrected mass shift was found to be 95~MeV at 
 $\sqrt{s}= 170$~GeV {\cite{tt10}}.
 The shift increases with increasing energy and
 decreasing source radius.

In the following analysis we will address the problem of
 Bose-Einstein correlations, but using another approach,
 which is based on assigning weights to the simulated events
 according to the momentum distributions of final state bosons.
 In the global event weight scheme a shift in the W mass can arise,
 if the event weight depends on \MW.
 The use of such global event weights is not straightforward and
 was discarded in the above study on fairly general grounds.
 Our aim is to reinvestigate the suitability of the method
 and to estimate approximately the systematic error on the W mass from 
 Bose-Einstein correlations. The use of global event weights is 
 complementary to the local reweighting scheme of ref.~\cite{tt10} in 
 the sense that here the kinematical properties
 of the events are preserved, while all probabilities and multiplicities
 may change. We have not found a unique solution to the problem of
 assigning the weights, however, and have used a number of different
 weighting schemes to assess the range of effects that can be
 expected from Bose-Einstein correlations.

 We have used the PYTHIA event generator {\cite{tt11}} to
 generate Monte Carlo samples of \ee\ $\to$ \WW\ $\to$ 4~jet events.
 A basic assumption here is that hadronic W and Z$^0$ boson decays
 are sufficiently similar, so that by using the tuning of the
 Monte-Carlo model parameters that reproduces the experimental
 data from Z$^0$ decays at LEP, Bose-Einstein effects in single
 W decays are already effectively taken into account in
 properties such as multiplicities and single particle momentum
 spectra.
 Hence, only the correlation between identical particles
 from different Ws have been included in the weight calculation. 
 In order to
 check the self-consistency and inherent systematic errors
 of the method we have applied the same
 weighting method to the well studied process of Z$^0$ hadronic decays.

 Since we want to estimate the maximal systematic error on the measured
 W mass due to the mass shift arising from Bose-Einstein  correlations,
 we have not taken into account smearing due to experimental resolution
 and acceptance, reconstruction method etc. 

 Various possibilities of constructing event weights
 are discussed in the following section.
 Some consequences 
 of these weighting schemes on certain measurable quantities 
 at the Z$^0$ peak are considered in Section 3, together 
 with possible constraints on BE effects from Z$^0$ physics. 
 In Section 4 the influence of BE effects
 on the process of W pair production at LEP2 is analyzed.

\section{Event weighting schemes for Bose-Einstein effects}

The Bose-Einstein effect corresponds to an enhancement
 in the production probability of identical bosons to
 be emitted with small relative momenta, as compared
 to non-identical particles under otherwise similar conditions.
 Experimental data are usually analysed in terms of the
 correlation function defined as the ratio of the two-particle
 probability density to the product of the corresponding single particle
 quantities:

\begin{equation}
 C(p_1,p_2) = \frac{P(p_1,p_2)}{P(p_1)P(p_2)}
\end{equation}
where $p_i$ is the four-momentum of particle $i$.
 Assuming a spherical space-time distribution of the particle source,
 the correlation function takes the form:

\begin{equation}\label{c1}
 C(Q) = 1 + \lambda \rho (Q)
\end{equation}
where $Q$ is the four-momentum difference, $Q^2 = -(p_1-p_2)^2$,
 and $\rho$ is equal to the absolute square of the Fourier transform 
 of the particle emitting source density, with the normalization condition
 $\rho(0)=1$.
 The parameter $\lambda$, known as the incoherence parameter, takes
 into account the fact that for various reasons, the strength of
 the correlations can be reduced. 

Often a gaussian model is assumed for the source density, which leads to
\begin{equation}\label{gauss}
 \rho (Q) = \exp(-R^2 Q^2)
\end{equation}
 where $R$ is the radius parameter. Experimentally, this gives a reasonable
 description of Bose-Einstein correlations in many types of collisions.
 In \ee\ collisions typical values of $R$ and $\lambda$ are 
 respectively 0.5 fm and 0.3 (1.0 for directly produced pions) 
 \cite{alephBE,delphiBE}.

In order to simulate the effects of Bose-Einstein correlations,
 we have in this study chosen to use the event weighting method.
 The method arises very naturally in a quantum mechanical approach, 
 where the weight can be constructed as the ratio of the square of the 
 symmetrized multiparticle amplitude to the square of the 
 non-symmetrized amplitude
 corresponding to the emission of distinguishable particles. 
 The use of global event weights leads
 to a number of conceptual and computational difficulties, however, which
 must be overcome for any quantitative conclusions to be drawn and which
 we will attempt to address below.
 
There are several possibilities to construct event weights.
 One way of forming the weight is to take a product of
 enhancements $C(Q)$ for all pairs of identical bosons
 in the event \cite{tt12}:

\begin{equation}\label{v1}
 V_1 = \prod_{{i_1,i_2}}C({Q_{i_{1}i_{2}}})
\end{equation}

For high multiplicity events this weight can become extremely large,
 so that a few such events dominate the weighted distributions and
 lead to non-realistic distributions. The event weights therefore
 have to be regularized in some way.
 In order to keep the statistical error at a reasonably
 low level, we have chosen to discard events with very high weights (higher
 than some $V_{max}$). This, however, has the unpleasant effect of 
 making the results $V_{max}$-dependent.
 We have tried to circumvent this difficulty by analyzing the 
 $V_{max}$ dependence 
 and extrapolating the results to $V_{max} \to \infty$.

One can also rescale the weight of the event using a single
 constant $w_0$:

\begin{equation}\label{v2}
 V_2 = V_1/w_0^n
\end{equation}
 where $w_0$ is a constant slightly larger than 1, and $n$ is the
 number of pairs in the event (i.e. the number of terms
 in the product in (\ref{v1})). The value of $w_0$ is chosen to 
 keep the average multiplicity reasonably close 
 to its value before event weighting {\cite{tt14}.}
 For a constant $V_{max}$, we have found, however, 
 that the method gives rise to numerical difficulties, stemming
 from the fact that increasing $w_0$ brings in more events from
 the high weight tail of $V_1$, which leads to large fluctuations in
 the multiplicity and the average event weight.
 Our results for the shifts in multiplicity and \MW\ using $V_2$ are
 roughly consistent with those found using $V_1$, and $V_2$ 
 will not be discussed further.

\begin{figure}
  \begin{center}
    \begin{picture}(100,100)(0,0)
    \epsfig{file=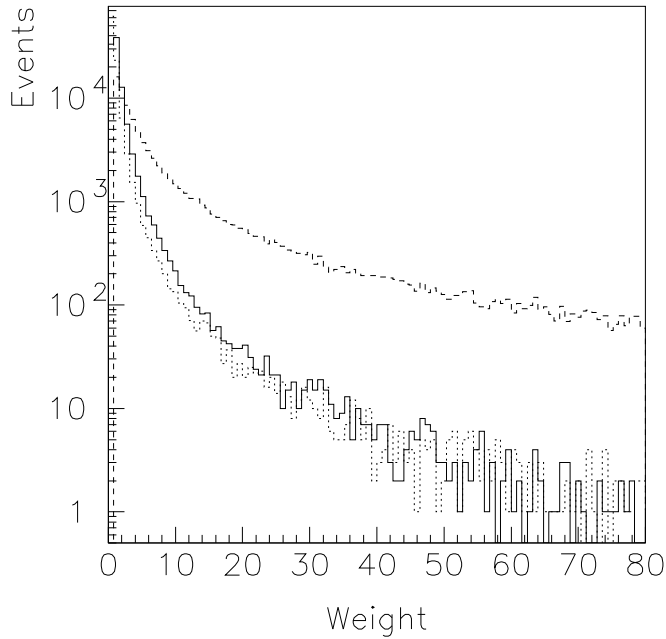,height=100mm}
    \end{picture}
  \end{center}
\caption{Distributions of the weights $V_1$ (dashed line), $V_3$ 
(full line) and $V_5$ (dotted line) for WW events at 175 GeV.}
\end{figure}

A problem with these weighting methods is that, like the local reweighting
 scheme of ref {\cite{tt10}}, they emphasize narrow jets and thereby introduce
 artificial correlations also between non-identical particles.
 In order to counteract this, one can use the weight calculated with 
 non-identical pairs to rescale (\ref{v1}):

\begin{equation}\label{v3}
 V_3 = V_1/V_0^{n/m}
\end{equation}
where $V_0$ is the weight calculated according to (\ref{v1}) but for
 non-identical bosons in the same event, while $n$ and $m$ are
 the numbers of identical and non-identical pairs, respectively.
 This also leads to a better numerical behaviour, 
 as illustrated in fig.\ 1, which shows
 the distributions of $V_1$ and $V_3$ for simulated WW events at 
 $\sqrt{s} = 175$ GeV. The high weight tail is much less
 pronounced for $V_3$ than for $V_1$.
 Both fall off to a good approximation as inverse powers,
 with exponents $-2.6$ and $-1.4$, respectively, which makes it plausible
 that the sum of all weights converges.

A different method of constructing the event weight, which is closer to
 a full quantum mechanical treatment, starts from the
 introduction of a symmetric amplitude, which has $n!$
 terms {\cite{tt14}}. This leads to a weight:

\begin{equation}\label{v4}
 V_4 = \sum_{{{\rm{permutations}}}} \lambda^{k/2}{\rho}(Q_{1i_1})
     {\rho}(Q_{2i_2}) \cdots {\rho}(Q_{ni_n}),
\end{equation}
where $k$ is the number of times when the first and
 second indices differ. For $Q=0$, $\lambda=1$
 and $n$ identical particles, equation (\ref{v1}) gives
 a weight of $2^{n(n-1)/2}$,
 while eq. (\ref{v4}) results in the correct  value $n!$.
 However, for typical hadronic configurations this difference
 is much smaller, and (\ref{v4}) is rarely used because of
 computational difficulties. We had to restrict ourselves to events
 containing no more than 8 identical particles of each kind.
 This limitation is too restrictive already at Z$^0$ energies,
 where the number of identical boson combinations is lower than in 
 \WW\ events, and rejected about 50 \% of events in W pair production, 
 making it
 effectively useless for the latter case. We used it only to check
 that it gave essentially consistent results with our other methods
 for events with low multiplicity.

The event weights defined above were based on the
 gaussian parametrization of the particle emitting source.
 This implies that $V_1 \geq 1$ for all values of $Q$.

This is not true in all models, however. In addition
 to the above weights, we have therefore also studied 
 a different pair weight, inspired by the weight used in \cite{bo},
 where $\rho$ in (\ref{c1})
 is not required to be always positive: 

\begin{equation}\label{wb}
 \rho(Q) = {{\cos (\xi Q R)}\over{\cosh (Q R)}}
\end{equation}

For $\xi$ close to 1, 
this is very close to (\ref{gauss}) apart from
 becoming slightly negative at large $Q$. The corresponding weight
 $V_5$ is built in analogy with (\ref{v1}), but with the gaussian 
 (\ref{gauss}) replaced by (\ref{wb}) (the dotted line in Figure 1). We
 find that $\xi=1.15$ 
 leads to a good overall description. Due to the better
 numerical behaviour of this weight function (the exponent in the
 power fit is $-2.4$), we were able to apply $V_5$
 without further rescaling.

\section{Influence of event weighting on Z$^0$ properties}

Various measurable properties of the Z$^0$ will be affected
 to different extent, if one introduces event weights into the
 simulation of its hadronic decays. Since the partonic states 
 before hadronization are known to be well described by
 perturbative calculations, which do not take into account
 Bose-Einstein correlations, uncritical application of
 event weights may lead to large inconsistencies with e.g.\
 measured branching ratios and relative frequencies of jet 
 multiplicities etc. In order to see how serious these effects
 are and to judge what consequences this has for the
 analysis of the WW events, the precise experimental
 data from Z$^0$ decays can be used to check the event
 weighting schemes of Bose-Einstein correlations for W pair
 production at LEP2.
 We have simulated $3\times 100000$
 hadronic events at $\sqrt{s}=M_{\mbox{Z}}$ and $M_{\mbox{Z}} \pm 2$~GeV.
 Table 1 presents the differences for charged particle multiplicity,
 shift of Z$^0$ peak position in hadronic vs leptonic decay modes,
 branching fractions for charm and beauty decays
 ($R_c$ and $R_b$) and the ratio of three to two jet events 
 with and without event weighting.

\begin{table}
\begin{center}
\begin{tabular}[tbc]{|l|rrr|} 
\hline
        & $V_1$  & $V_3$  & $V_5$  \\
\hline
$\Delta \langle n_{ch} \rangle$   & $3.7 \pm 0.5$
       & $1.3 \pm 0.2$ & $1.8 \pm 0.2$ \\

$\Delta M_{Z_0}$,~MeV & $8\pm 3$ & $  0\pm 3$ &      $ 1\pm 4$ \\

$\Delta R_c,$ \%      & $-3\pm 2$ & $  -2\pm 2$ &      $0\pm 2$ \\

$\Delta R_b,$ \%      & $-26\pm 3$ & $-11\pm 2$ &      $-5\pm 2$ \\

$\Delta$ 3jet/2jet, \%      & $80\pm 20$ & $  20\pm 5$ &  $ 20\pm 5$ \\
\hline
\end{tabular}
\end{center}
\caption{
 Differences in charged multiplicity, peak mass of Z$^0$, branching
 fractions and three-to-two jet event ratio, between weighted
 and non-weighted events, for various weighting systems described
 in the text. 
}
\end{table}

This analysis resulted in the following: 
\begin{itemize}
\item[-]
 The average charged multiplicity has changed. The weight
 $V_1$, which was not rescaled, leads to the largest
 increase when compared to the unweighted results, while both $V_3$ 
 and $V_5$ give a smaller increase around 1.5. In all these cases, 
 the change can be accommodated by
 retuning of the parameters in the simulating program.
\item[-]
 In principle, event weighting can result in a shift of 
 the Z mass peak.
 However, only $V_1$ yielded a shift of a few MeV, while for 
 $V_3$ and $V_5$ the shift is essentially zero.
 We have not found any significant change of the Z$^0$ width.
\item[-]
 The pattern of heavy and light quark fragmentation is rather different.
 Heavy quarks produce significantly less pairs with small $Q$, and all BE
 effects in this approach are less pronounced for heavy
 quarks. Heavy quark events thus obtain smaller average weights, which
 result in changes shown in Table 1. Note that the effect for $c$-quarks
 is diluted because $b$-quark events reduce the overall average 
 weight.
 Further study of this effect lies beyond the scope of this work. In
 order to exclude this artificial 
 flavour dependence in W decays, the weighting
 and rescaling was performed separately for the different decay modes of the
 Ws.
\item[-]
 The weighting resulted in a
 substantial increase of jet activity, as measured by the three to
 two jet event ratio. This is however difficult to
 quantify because of its dependence upon the jet finding algorithm and its
 parameters. The numbers shown in Table 1 were obtained using
 LUCLUS with default parameters, corresponding to fairly 
 narrow jets. The effect decreases for broader jets and in any case is 
 much less pronounced in WW production, so we did not attempt to
 correct for it.

\end{itemize}

We have considered maximum BE correlations,
 $\lambda =1$, but only for pions and kaons originating from the sources
 with decay lengths $c\tau < 10$~fm, and $\lambda = 0$ otherwise.
 The source radius $R$ was taken to be equal to 0.5~fm everywhere.
 In Z$^0$ decays at LEP1 one observes $\lambda \approx 0.3$ if all
 particles are considered, 0.4 if only pions are taken into account
 and 1.0 for directly produced pions {\cite{alephBE,delphiBE}}, 
 and $R \approx 0.5$~fm.

The reproduced correlation functions for the three weighting schemes, 
 $V_1$, $V_3$ and $V_5$ are shown in figure 2.  
 We have divided the ratio of the weighted same-sign particle
 distribution to the unweighted one by the similar ratio for
 opposite-sign particles. This is equivalent to using the 
 $Q$-distribution of opposite-sign pairs as a reference sample and 
 correcting for effects of particle selection and resonances by dividing
 by the same ratio in simulated events without BE correlations --- 
 a common procedure in experimental analyses. 
 Also shown are fits to the form
 \begin{equation}
   N(1 + \beta Q)(1 + \lambda \exp (-Q^{2}R^{2}))
 \end{equation}
 which is often used to parametrize the experimentally observed correlation
 function in Z$^0$ decays \cite{alephBE,delphiBE}. 
 The resulting values of the parameters for a fit range of 
 0--2 GeV in $Q$ are shown in Table 2. The dashed
 line in the figure represents the result of a fit to the correlation function
 of all particles observed in real data from hadronic 
 Z$^0$ decays {\cite{delphiBE}}.

\begin{figure}[tb]
  \begin{center}
    \begin{picture}(140,100)(0,0)
    \epsfig{file=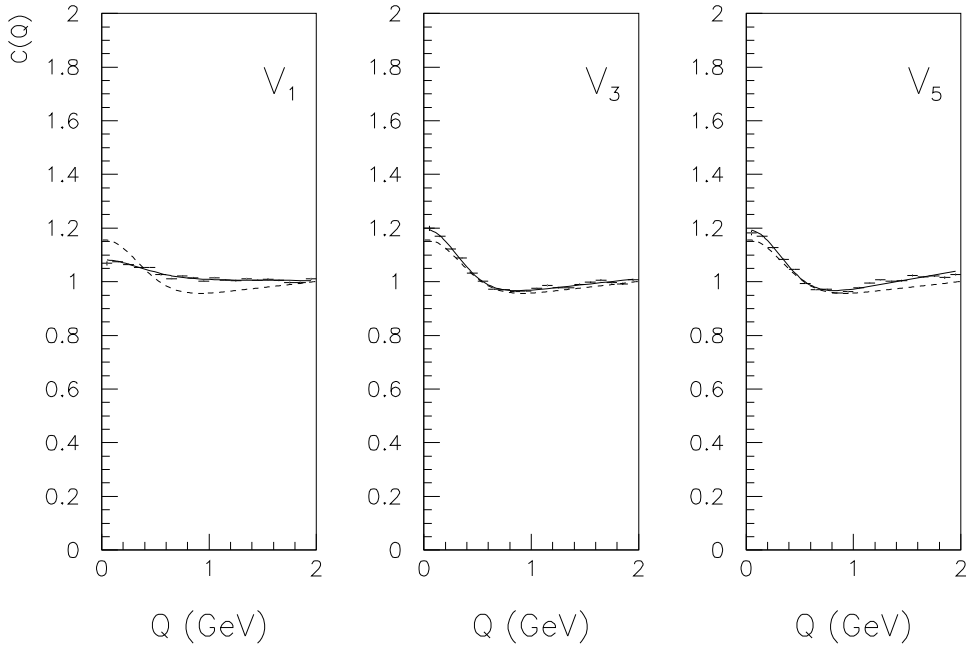,height=100mm}
    \end{picture}
  \end{center}
\caption{Reproduced correlation functions in Z$^0$ events using the 
weights $V_1$, $V_3$ and $V_5$. The dashed lines show the result of a 
fit to real data from hadronic Z$^0$ decays
\protect\cite{delphiBE}.}
\end{figure}
$V_1$ does not reproduce the correlation function well, as it gives
 too small values for both the incoherence parameter $\lambda$ and the
 radius $R$. The other weight schemes give very reasonable 
 descriptions resulting in input and output parameter values which are 
 reasonably close to each other. We take the
 spread of the results using the different weighting schemes to be
 indicative of the systematic errors inherent in the method. 

\begin{table}
\begin{center}

\begin{tabular}[tbc]{|l|rrr|}
\hline
 Weight   &   $V_1$         &     $V_3$       &    $V_5$          \\
\hline
$\lambda$ & $0.072\pm .004$ & $0.292\pm .005$ & $0.325\pm .005$   \\
\hline
$R$~(fm)  & $0.387\pm .018$ & $0.465\pm .005$ & $0.455\pm .005$   \\
\hline
$\beta$ (GeV$^{-1}$) & $-0.029 \pm .003$ & $0.047 \pm .003$ 
                                              & $0.080 \pm .003$  \\
\hline
$N$       & $1.010\pm .004$ & $0.925\pm .004$ & $0.899\pm .004$   \\
\hline
\end{tabular}
\end{center}

\caption{
 The fitted values of the correlation function parameters
 $\lambda$, $R$, $\beta$ and $N$ at the Z$^0$ peak.
}
\end{table}

Hence we conclude that, provided that the different quark final states
 (and possibly the final states with different number of jets)
 are treated separately, application of the global event weighting
 technique with rescaling of the weight ($V_3$) or using the form ({\ref{wb}})
 is not inconsistent with LEP1 data at the Z$^0$, whereas the direct
 application of the product of pair weights ($V_1$) should be treated with
 more care.

\section{W pair production}

PYTHIA 5.7 was used to simulate the process \ee\ $\to$ \WW\ $\to$ 4~jets,
 and the weighting schemes described above were applied to generate BE
 effects. 

Only correlations between identical bosons originating
 from different Ws were included, since BE correlations
 within a single W cannot lead to any 
 change in \MW\ compared to the semileptonic
 decays  \ee\ $\to$ \WW\ $\to 2$~jets$+l\nu$. 

The first thing to study is the correlation function and the compatibility
 of input and output values indicating the self-consistence of the method.
 Table 3 refers to 175~GeV, and contains
 the results of the fits to the correlation functions for each weight
 used (as above for Z$^0$) using the cut off $V_{max}=80$.
 Numbers for 192~GeV are very similar.
 The errors given in the table are statistical only. From the variation of 
 the fit results with $V_{max}$, we estimate that the
 systematic error is 0.05 on $\lambda$ and 0.05 fm on $R$.

\begin{table}
\begin{center}
\begin{tabular}[tbc]{|l|rrr|}
\hline
 Weight   &      $V_1$      &    $V_3$        &  $V_5$          \\
\hline
$\lambda$ & $0.032\pm .004$ & $0.101\pm .002$ & $0.146\pm .003$ \\
\hline
$R$~(fm)  & $0.322\pm .024$ & $0.459\pm .009$ & $0.457\pm .006$ \\
\hline
$\beta$ (GeV$^{-1}$) & $0.002\pm .003$ & $0.013\pm .001$ 
                                             & $0.036\pm .001$ \\
\hline
$N$       & $0.997\pm .003$ & $0.977\pm .003$ & $0.949\pm .003$ \\
\hline
\end{tabular}
\end{center}

\caption{
The fitted values of the correlation function parameters
 $\lambda$, $R$, $\beta$ and $N$ for W pair production at 
 $\protect\sqrt{s} = 175$ GeV.
}
\end{table}

One sees that for $V_1$, $R$ is somewhat lower that the
 input value, while $V_3$ and $V_5$ give values quite close to 0.5.
 It is worth noting that the values of $\lambda$ are significantly
 smaller than at the Z peak, essentially because in the WW case we have
 included only correlations between pairs from different
 Ws.

Next, the information from the Monte-Carlo was
 used to assign each final particle to the W$^+$ or the W$^-$,
 as in {\cite{tt10}}.
 Ws with the mass values in the interval
 $70~{{\rm GeV}} \leq M_{W} \leq 90$~GeV
 were  studied at 175 and 192~GeV to assess the
 energy dependence. At each energy, $10^5$ events were generated,
 which is about an order of magnitude higher than the
 expected statistics of all four LEP experiments combined at
 500~pb$^{-1}$ integrated luminosity per experiment.
 In general, one expects that BE-induced effects in WW production should
 die out at high energies, as the overlap between the two W decay
 volumes decreases. This requires much higher energies than will become
 available at LEP2, however, and it is likely that
 the effect will increase with energy in the LEP2 range \cite{tt10}.

The mass distribution of W bosons was built with and
 without event weighting for each of the weights used,
 and the differences were calculated
 in the average charged multiplicity $n_{ch}$,
 the mean W mass, $M^{\rm{mean}}_W$,
 averaged over the whole interval $70~{{\rm GeV}} \leq M_{W} \leq 90$~GeV,
 and a fitted $M^{\rm{fit}}_W$.
 The fit was performed using a relativistic Breit-Wigner shape with an
 $s$-dependent width, in the interval $ 80.25\pm \delta$~GeV, with
 $\delta = 2$ GeV. The results are presented in Table 4.
 We also performed the fit in wider intervals corresponding to
 $\delta = $ 4,6,8 and 10 GeV and observed a small additional increase in the
 mass shift of a few MeV, which saturated at the larger interval sizes,
 consistent with the fact that the tails of the Breit-Wigner distribution
 contain very little information about the value of the W mass.

As mentioned above, for computational reasons, we were forced to cut
 away a tail of events with very large weights.
 We have tried to eliminate
 the dependence on the cutoff value, $V_{max}$, by calculating 
 the multiplicity and mass shifts for
 three values of $V_{max}$ (20, 40 and 80) and then extrapolating to
 infinite cutoff. This method seems to be more reliable and less
 vulnerable to fluctuations than direct calculation with very high
 $V_{max}$.
 Figure 3 shows the values of the mean W mass, $M^{\rm{mean}}_W$,
 and the fitted $M^{\rm{fit}}_W$ as functions of $1/V_{max}$. 
 There is no indication in our investigations that the inclusion of the
 events with very large weights would change the estimated mass shifts
 by any significant amounts. For $V_1$, the extrapolated value depends
 on the specific way the extrapolation is performed. We have included this
 ambiguity into the error shown in Table 4.

\begin{figure}[tb]
  \begin{center}
    \begin{picture}(140,100)(0,0)
    \epsfig{file=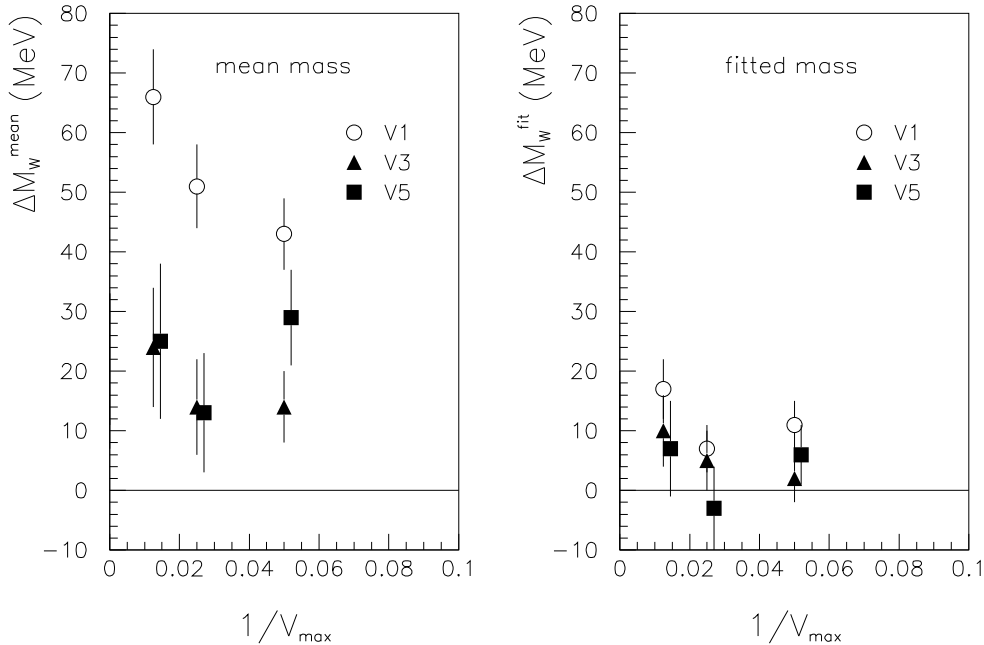,height=100mm}
    \end{picture}
  \end{center}
\caption{
Shifts in the mean and fitted W mass as functions of
the inverse weight cut off, $1/V_{max}$, for weight schemes $V_1$, $V_3$ 
and $V_5$, and $\protect\sqrt{s} = 175$ GeV.
}
\end{figure}

\begin{table}
\begin{center}

\begin{tabular}[tbc]{|l|rrr|}
\hline
                & $V_1$  & $V_3$  & $V_5$  \\
\hline
$\Delta n_{ch}$    &          &          &        \\

 175 GeV           &  $3.8 \pm 0.5$ &  $1.8 \pm 0.2$  & $ 1.0 \pm 0.2$ \\
 192 GeV           &  $3.7 \pm 0.5$ &  $1.7 \pm 0.2$  & $ 0.6 \pm 0.2$ \\
\hline

$\Delta M_W^{mean}$~(MeV)  &     &          &        \\

  175 GeV          &  $75 \pm 15$  &  $22 \pm 11$  &  $20 \pm 14$ \\
  192 GeV          &  $92 \pm 16$  &  $34 \pm 11$  &  $38 \pm 14$ \\
\hline

$\Delta M_W^{fit}$~(MeV) &       &          &       \\

 175 GeV          &  $12 \pm  9$  &  $11\pm 7$    &  $ 4 \pm 12$ \\
 192 GeV          &  $15 \pm  8$  &  $13\pm 7$    &   $6 \pm  9$ \\
\hline
\end{tabular}
\end{center}

\caption{
Values of differences in multiplicity and mass of the W boson for events
 with and without interconnecting Bose-Einstein correlations between the
 two Ws. 
}
\end{table}

From these numbers we draw the following conclusions:

\begin{itemize}

\item[-]
There is a clear correlation between the BE-induced shifts in the W mass
 and in the charged particle multiplicity in WW production: the larger
 is the increase in charged multiplicity, the larger mass shifts are
 expected.

\item[-]
Both $V_3$ and $V_5$ weights result in fairly
 small shifts, while still maintaining a good reproduction of 
 the correlation function. They are well-behaved numerically 
 and in our opinion give quite reliable estimates of the effect.
 However, we conservatively take
 the spread of values using all three weighting schemes as indicative of
 the systematic errors inherent in our approach. 

\item[-]
The fitted value for the W mass is less sensitive to BE effects
 than the mean over the full distribution, which has been used
 to estimate the effect in previous investigations \cite{tt10}.
 Our estimated values for the shift in the fitted mass are
 less than 20 MeV, implying that BE correlations are not too
 dangerous for the W mass measurements at the expected level of accuracy
 at LEP2. For the shifts in the mean W mass, we confirm the conclusions
 of {\cite{tt10}} and find values of the same general magnitude of a few
 tens of MeV, under
 similar conditions as investigated there. In all cases the shift is 
 towards larger masses, as expected on general grounds \cite{tt7,tt10}.

\item[-]
For all weighting schemes, the shift in \MW\ increases
 with energy in the energy range considered, but the increase is fairly
 small.

\end{itemize}

It is interesting to compare our  results to the
 predictions based on the implementation of Bose-Einstein
 effects by shifting the momenta of final state particles
 {\cite{tt10}}. There are several differences in the predictions
 of these two schemes.

 The most important difference is in the particle multiplicity:
 our approach naturally leads to an increase of the average number of
 particles due to Bose-Einstein correlations, while the
 momentum-shifting method assumes that the multiplicity is
 unchanged. Experimentally, it is not known yet to what extent Bose-Einstein
 correlations might modify the particle multiplicities at high energies.

 The energy dependence of W mass shift is different.
 The strong energy dependence in  momentum-shifting scheme
 is a combination of two effects: the increase of the systematic
 shift for low momentum particles in the direction of smaller
 W momenta, and the differences in momentum spectra of
 W decay products for various energies, as stressed
 in {\cite{tt10}}. This seems to be less pronounced in the present
 approach.

The present study confirms that the systematic effect of BE correlations
 on the W mass determination can potentially be quite large, as found
 in \cite{tt10}, although the actual values of the
 mass shift found here are somewhat smaller. The size of the shift is
 however quite sensitive to the procedure used to extract the value of the
 W mass. In particular we observe that a fit to the
 lineshape of the W mass distribution has a much smaller systematic error 
 from Bose-Einstein correlations than the average mass, due to the
 fact that the main effect on \MW\ in our scheme arises from the tails
 of the mass distribution, which contain very little information about
 the peak position. Hence it does seem possible to keep the systematic 
 error from this source below about 20 MeV. Careful work linked 
 to the actual fitting procedures used
 by the LEP experiments is obviously needed in order to assess this in 
 the individual cases and to optimize the analysis procedures.
 Since the value of the mass shift
 is always positive (as also expected on general grounds), a further
 reduction of the systematic error by a factor two is in principle
 possible by assigning the expected shift as a correction to \MW.

The estimates in this paper were all made assuming full Bose-Einstein
 correlation strength ($\lambda= 1$). If this should turn out not to be the
 case experimentally (see \cite{delphiBEWW}), the effect on the W mass 
 may be correspondingly reduced. Here we just want to bring attention 
 to the fact, that such a reduction cannot a priori be expected to depend 
 proportionally on $\lambda$, as it is presumably not the correlation 
strength, 
 but the absolute number of correlated pairs of pions in the enhanced region of 
 the $Q$ distribution that is important for the size of the possible 
 shift in the W mass.

The comparison of hadronic decays of
 $\mbox{W}^+\mbox{W}^- \rightarrow q\bar qq \bar q$ and
 $\mbox{W}^+\mbox{W}^- \rightarrow q\bar ql\nu$
 channels gives a unique possibility to investigate the
 influence of Bose-Einstein correlations on various properties
 of final state particles, such as multiplicity, transverse
 and longitudinal momentum spectra, resonance properties and reconstructed
 jet characteristics. It is possible, that by taking proper care
 in the fitting procedures used, one may at the
 same time be able to use a large part of the hadronic \WW\ events for the
 W mass determination, and to study the interconnection
 effects in the relatively clean setting of \ee\ $\to$ WW events, 
 by restricting the study to the region of large, off-peak W masses 
 where these effects are expected to be the largest.

We would like to thank A.~Olshevsky, S.A.~Gogilidze, R.~Lednicky,
 V.L.~Lyuboshitz, T.~Sj\"{o}strand and A.~Tomaradze for helpful discussions.
 One of us (V.K.) would like to thank ALEPH collaboration for their kind
 hospitality at CERN, where a part of this work was done.

{\bf{Note added.}} Results of a similar study have been published recently
{\cite{jad}}. The weight system used in \cite{jad} is a simplified version
 of our $V_4$, where the computational difficulties were avoided by
 averaging over ``clusters'' of particles. The authors do not see any mass
 shift due to BE correlations at the level of their statistical precision.
 This is not inconsistent with the present calculation, as they have 
 chosen to use $R=1$~fm, which strongly reduces the number of interfering
 pairs.


\begin{thebibliography}{99}

\bibitem{tt1} J.~Alitti et al., (UA2 Collab.), Phys.\ Lett.\ B276 (1992) 354.
\bibitem{tt2} F.~Abe et al., (CDF Collab.), 
 Phys.\ Rev.\ Lett.\ 65 (1990) 2243;
 Phys.\ Rev.\ D43 (1991) 2070;
 Phys.\ Rev.\ Lett.\ 75 (1995) 11.
\bibitem{tt3} S.~Abachi et al., (D0 Collab.), FERMILAB-PUB-96-177-E.
\bibitem{pdg} R.M.~Barnett et al., Review of Particle Physics,
 Phys.\ Rev.\ D54 (1996) 1.
\bibitem{tt4} P.B.~Renton, in proc.\ of the XVII Int.\
 Symp.\ on Lepton-Photon Interactions, August 1995, Beijing, China, p.~35.
\bibitem{tt5} J.~Bijnens et al., ECFA Workshop on LEP200,
 CERN 87-08 (1987), Vol.\ I, 49.
\bibitem{tt6} G.~Altarelli et al., The Workshop on Physics at LEP2,
 CERN-TH/95-151.
\bibitem{tt7} Z.~Kunszt et al., Physics at LEP2,
 CERN 96-01 (1996), Vol.\ I, 141.
\bibitem{opalww} K.~Ackerstaff et al., (OPAL Collab.), CERN-PPE/96-141;\\
 P.~Abreu et al., (DELPHI COllab.), CERN-PPE/97-09;\\
 M.~Acciari et al., (L3 Collab.), CERN-PPE/97-14;\\
 R.~Barate et al., (ALEPH Collab.), CERN-PPE/97-25.
\bibitem{lepww} LEP EW Working group, LEPEWWG/97-01.
\bibitem{tt8} V.~Kartvelishvili and R.~Kvatadze, MC-TH-95/18, 1995.
\bibitem{tt9} G.~Gustafson et al., Phys. Lett., B209 (1988) 90.
\bibitem{tt0} T.~Sj\"{o}strand and V.~Khoze, 
                   Phys.\ Rev.\ Lett.\ 72 (1994) 28;\\
 T.~Sj\"{o}strand and V.~Khoze, Z.~Phys.\ C 62 (1994) 281.
\bibitem{tt10} L.~L\"{o}nnblad and T.~Sj\"{o}strand, 
 Phys. Lett., B351 (1995) 293.
\bibitem{tt11} T.~Sj\"{o}strand, Computer Physics Comm.\ 82 (1994) 74.
\bibitem{tt12} G.D.~Lafferty, Z.~Phys.\ C60 (1993) 659.
\bibitem{tt13} J.E.G.~Edwards and G.D.~Lafferty, MAN/HEP/94/6, 1994.
\bibitem{alephBE}D.~Decamp et al., (ALEPH Coll.), Z.~Phys.\ C54 (1992) 75.
\bibitem{delphiBE}P.~Abreu et al., (DELPHI Coll.), 
 Phys.~Lett.\ B286 (1992) 201; Z.~Phys.\ C63 (1994) 17.
\bibitem{tt14} S.~Haywood, RAL-94-074, 1994.
\bibitem{bo} B.~Andersson, in Proc.\ XXV Int.\ Symp.\ on Multiparticle
 Dynamics, WS, 1995, p. 335; B.~Andersson and M.~Ringner, LU TP 97-07,
 April 1997, hep-ph/9704383.
\bibitem{delphiBEWW} P.~Abreu et al.,(DELPHI Collab.) CERN-PPE/97-30.
\bibitem{jad} S.~Jadach and K.~Zalewski, CERN-TH/97-29, February 1997.

\end{thebibliography}
\end{document}